\documentclass[aps,prb,twocolumn,english,showpacs,letterpaper,superscriptaddress]{revtex4-1}

\usepackage[dvips]{epsfig}
\usepackage[colorlinks=true,citecolor=blue,linkcolor=blue]{hyperref}

\usepackage{amssymb}
\usepackage{graphicx}
\usepackage{subfigure}
\usepackage{array}

\usepackage{epsfig}

\usepackage{amsmath}
\usepackage{color}
\usepackage{float}
\usepackage{mathrsfs}
\usepackage{indentfirst}
\usepackage{textcomp}
\usepackage{comment}
\usepackage{mathtools}

\newcommand{\ignore}[1]{}
\newcommand{\mater}{ZrTe$_5$}

\newcommand{\titledef}{Robust and tunable Weyl phases by coherent infrared phonons in \mater} 

\begin{document}
\bibliographystyle{naturemag}

\title{\titledef}

\author{Niraj Aryal}
\email{naryal@bnl.gov}
\affiliation{Condensed Matter Physics and Materials Science Division, Brookhaven National Laboratory, Upton, New York 11973, USA}
\author{Xilian Jin}
\affiliation{Condensed Matter Physics and Materials Science Division, Brookhaven National Laboratory, Upton, New York 11973, USA}
\affiliation{State Key Laboratory of Superhard Materials, College of Physics, Jilin University, Changchun 130012, China}
\author{Qiang Li}
\affiliation{Condensed Matter Physics and Materials Science Division, Brookhaven National Laboratory, Upton, New York 11973, USA}
\affiliation{Department of Physics, Stony Brook University, Stony Brook, New York 11794, USA}
\author{Mengkun Liu}
\affiliation{Department of Physics, Stony Brook University, Stony Brook, New York 11794, USA}
\author{A. M. Tsvelik}
\affiliation{Condensed Matter Physics and Materials Science Division, Brookhaven National Laboratory, Upton, New York 11973, USA}
\author{Weiguo Yin}
\email{wyin@bnl.gov}
\affiliation{Condensed Matter Physics and Materials Science Division, Brookhaven National Laboratory, Upton, New York 11973, USA}

\date{\today}

\begin{abstract}
Ultrafast optical control of the structural and electronic properties of various quantum materials has recently sparked great interest. In particular, photoinduced quantum phase transition between distinct topological phases has been considered as a promising route to realize ultrafast topological quantum computers. Here we use first-principles and effective Hamiltonian methods to show that in \mater, a layered topological material, lattice distortions corresponding to all three types of zone-center infrared optical phonon modes can drive the system from the strong or weak topological insulating phase to a Weyl semimetal by breaking the global inversion symmetry.
Thus achieved Weyl phases are robust, highly tunable 
and one of the cleanest ones due to the proximity of the Weyl points to the Fermi level and a lack of other carriers.
We further show that the amount of infrared-mode pumping necessary to induce such Weyl phases can be reduced if used in conjunction with an A$_g$ Raman-mode pumping that first drives the system to the Dirac semimetal state.
We also find that Berry curvature dipole moment (BCDM), induced by the dynamical inversion symmetry breaking, gives rise to various nonlinear effects that oscillate with the amplitude of the phonon modes. \ignore{For the lowest-energy infrared phonon mode, BCDM changes sign as a function of the chemical potential only when the system is in the Weyl phase, which suggests an experimental way to detect the photoinduced Dirac-Weyl topological phase transition in \mater.}
These nonlinear effects present a novel switch for controlling the Weyltronics enabled quantum system.

\end{abstract}

\maketitle
\section{Introduction}
Topological materials, such as Dirac and Weyl semimetals, and topological insulators, have attracted monumental research efforts thanks to their novel properties and a potential for energy and quantum information applications~\cite{RevModPhysHasanKane2010,Armitage_Vishwanath_RMP2018}. 
Despite profound challenges, quantum information has become a major thrust in this field due to the robustness of the topological states~\cite{proximityeffectFuKane2008,JosephsonSupercurrentHgTe_Wiedenmann_Naturecomm2016}, highly desirable for high-temperature fault-tolerant multi-qubit computation and communication ~\cite{TopoQuantComp_Nayak_Das_RMP2008,ChiralQubit_Kharzeev_Li_2019,TopoQuantChiralMajorana_Lian_Zhang_PNAS2018}. The fast operation on the qubits requires a quantum material that can be easily driven from one topological phase to another by small controllable stimuli, in particular by the ultrafast laser pumping ~\cite{UltrafastSymmetrySwitchWeyl_SieNature2019,SymmetrySwitchMoTe2_ZhangPRX2019,Weber_JAP_21_review}. 
It would be highly desirable to have  a clean system that unambiguously displays signatures of the topological characters in different transport phenomena.


 \ignore{
Condensed matter realizations of the Weyl fermions, which require broken inversion or time reversal symmetry, 
are one of the most robust ones and have been posed as one of the viable alternatives for topological qubits~\red{cite Q. Li's paper on Chiral qubits}.
These fermions do not require any other symmetry for their protection, except translational one, 
and give rise to various novel types of transport ~\cite{Armitage_Vishwanath_RMP2018,TransportDiracWeylReview_Wang_Liao_AdvancesInPhysics2017}.
Despite a great amount of  Weyl materials have been discovered ~\cite{WeylPyrochloreVishwanath2011,WeylTaAsHuang2015,TypeIISoluyanov2015}; the search for tunable and clean Weyl systems containing the minimal number of Weyl points in the vicinity of the Fermi surface
without trivial carriers continues~\cite{TopologicalIntercalated_Inoshita_Murakami_PRB2019,ScanWeyl_Xu_Sun_NPJ2020}.
}

Zirconium pentatelluride ({\mater}) is a tunable topological material ideally suited for studying different topological phases due to its proximity to the phase boundary between the weak topological insulator (WTI) and the strong topological insulator (STI)~\cite{QSHZrTe5_PRX_Weng_Dai_2014,CME_QiangLi_Nature2016,ZrTeARPES_Manzoni_PRL2016,ZrTeARPES_WeakTI_XiongPRB2017}. 
Owing to the small band gap ($\sim$ 20 meV) and layered geometry, small external perturbations such as strain and temperature can drive this system from the STI to WTI regime with Dirac semimetal (DSM) as a critical point~\cite{StrainTunedTopology_Mutch_Science2019,TempDrivenTopology_Xu_PRL2018}. By breaking time reversal symmetry, magnetic fields of a few Tesla were predicted to change the system to Weyl and nodal line semimetal depending on the direction of the field~\cite{ChenEffectiveH_PRL_2015} and a chiral
magnetic effect in electron transport was observed~\cite{CME_QiangLi_Nature2016}.

Recently, it was shown that  the STI-DSM-WTI transition  
can be achieved in \mater~by photoexciting various combinations of the Raman phonon modes~\cite{aryal2020topological,Konstantinova2020photoinduced,RamanZrTe5_VaswaniPRX2020}, which preserves the inversion symmetry. Since  4$\times$4 Dirac Hamiltonian (massive or massless) can become 2$\times$2 Weyl Hamiltonians separated in $\bf{k}$-space if either time reversal or inversion symmetry is broken, this suggests  that it might be possible to achieve a Weyl semimetallic phase by breaking inversion symmetry~\cite{3DGapless_Murakami_NJP2007,Weyl_Balents_PRB2012} with infrared (IR) phonon modes in \mater~\cite{aryal2020topological,wang2020expansive}.
Luo \textit{et. al}~\cite{Luo_NP_21_ZrTe5} have demonstrated that nonlinear photocurrent and chiral charge pumping in {\mater} could be generated by using a circularly polarized high-intensity (with fluence of $\sim$1 mJ/cm$^{2}$) 800nm laser source that appeared to induce the lowest IR phonon mode. While bulk photocurrent generation in inversion asymmetric crystals has been known for quite some time~\cite{CircularPhotogalvanic_Asnin_SSC_1979,Photogalvanic_Belinicher_IOP_1980}, the above experimental observation manifests photocurrent generation in an inversion symmetric material.
Its significance in light controlled quantum topological switches makes it urgent a thorough investigation of the IR phonon induced topological phase transition and related transport signatures in \mater. For example, it is not obvious if the photocurrent generation process and the topological phase transition to the Weyl semimetallic phase reported in Ref.~\onlinecite{Luo_NP_21_ZrTe5} is specific to a particular IR phonon mode or a general behavior induced by all (or several different combinations) of the IR modes due to the breaking of the underlying crystal inversion symmetry. 
The later possibility will be particularly useful for photocurrent engineering and optimisation.  
Another key issue is the high cost of the high intensity (or fluence) laser sources used to drive dynamical inversion asymmetry in such ultrafast experiments. 
Hence it is important to figure out whether it is possible to generate photocurrent as well as drive the system to the topological phase transition by a smaller lattice distortion.  
Finally, it is intriguing to explore the role of the Fermi surface and the Weyl points (WPs) in the second order nonlinear phenomena such as photocurrent and non-linear Hall effect.

In this article, we address these questions by performing a systematic computational and theoretical study of ZrTe$_5$ under
adiabatic atomic displacement corresponding to the IR optical phonon modes. Time-dependent density functional theory calculations~\cite{Huebener_NC_17_TDDFT} showed how femtosecond laser pulses with circularly polarized light can turn the Dirac semimetal into the Weyl semimetal in Na$_3$Bi, a symmetry-protected three-dimensional Dirac semimetal~\cite{Wang_PRB_12_Na3Bi,Liu_Science_14_Na3Bi}, while linearly polarized pumping can induce a symmetry breaking field that destroys the Dirac point and opens a gap. The further justification for our approach comes from the ultrafast experiments which have demonstrated the existence of a long-living (100-200 ps) photoinduced transient electronic state in \mater \cite{RamanZrTe5_VaswaniPRX2020,Luo_NP_21_ZrTe5}, accompanied by a similarly long-living shifting of the atomic positions \cite{Konstantinova2020photoinduced}. This long thermalization time is consistent with the low electronic density of the states at the tip of the Dirac cone. 


 We have found that
i) atomic displacements corresponding to all possible types of $\Gamma$-centered IR phonon modes in ~\mater~ introduce inversion asymmetry in only three possible ways; all these three types of IR modes dynamically drive the system from STI to the Weyl phase,
ii) these dynamical Weyl phases exist only when the magnitude of the normal atomic displacements (or equivalently lattice  distortion) exceeds  a certain threshold which magnitude depends on the details of the phonon modes; then such Weyl phases are robust over a large value of the lattice distortion and are highly tunable in the number and position, 
iii) the magnitude of lattice distortion, which is determined by the fluence of the laser source used in the experiments, can be dramatically reduced if the system is in close vicinity to the Dirac semimetallic (DSM) phase, and 
iv) while the phonon-driven inversion asymmetry naturally leads to the nonlinear responses induced by the non-zero Berry curvature, the occurrence of the WPs in the vicinity of the Fermi surface significantly changes their magnitude and direction.
Such tunable non-linear responses may find applications in various quantum switches~\cite{BulkRectification_Ideue_NaturePhys2017}.   
We believe that the proximity of the dynamical WPs to the Fermi level and  lack of other trivial carriers make the phonon-driven \mater~  
an ideal platform for understanding and verification of many intriguing physical properties often attributed to the presence of the WPs. 


\ignore{
The organization of this article is as follows.
In Sec.~\ref{Sec:Computation}, we present the details of our computational methods.
Then in Sec.~\ref{Sec:Results}, we present our results and discuss them in detail.
In Sec.~\ref{Sec:Discussion}, we present our discuss our results and their implications to the past and future experiments.
Finally, in Sec.~\ref{Sec:Conclusion}, we present our conclusion
and future outlook.
}

\section{Results}
\label{Sec:Results}
\subsection{Crystal structure and IR phonon modes}
\ignore{
\begin{figure}[htb]
            \includegraphics[width=0.48\textwidth]{3IrModes.jpg}
            \caption{Three different types of IR phonon modes of \mater~ projected onto the $\mathbf{b}-\mathbf{c}$ plane with the vectors showing the normal atomic displacement: (a) B$_{1u}$-4 (b) B$_{2u}$-20 and (c) B$_{3u}$-8 modes with frequencies of 0.58, 3.04 and 1.38 THz respectively. 
The atomic displacements in panel (c) are perpendicular to the $\mathbf{b}-\mathbf{c}$ plane, 
hence the vector are not seen in the figure. 
            }
        \label{fig:3IRModes} 
\end{figure}
}

\mater~ crystallizes in orthorhombic Cmcm space group. 
The primitive unit cell contains 2 formula units and thus $N_{\mathrm{atoms}}$=12. The calculated phonon band structure was reported in Ref.~\onlinecite{aryal2020topological}.
The 36 phonon modes of \mater~  
at the $\Gamma$-point can be written into the following 
irreducible representations of the isomorphic point group $D_{2h}$~\cite{Zwick_PhononsZrTe5_1982}:
\begin{eqnarray}
    \Gamma_{\mathrm{acoustic}} &=& B_{1u} + B_{2u} + B_{3u},\nonumber \\
    \Gamma_{\mathrm{optical}} &=& 6A_g + 2A_u + 4B_{1g} + 5B_{1u} + 2B_{2g} + \nonumber \\ && 5B_{2u} + 6B_{3g} + 3B_{3u},
\end{eqnarray}
where $g$ $\&$ $u$ stand for Raman and IR modes respectively and A $\&$ B modes denote symmetry and anti-symmetry with respect to the 2-fold symmetry axes.
The A$_u$ modes are optically inactive.
The phonons at the $\Gamma$-point can further be divided into groups of modes 
perpendicular and parallel to the chain direction i.e. \textbf{a}-axis~\cite{Zwick_PhononsZrTe5_1982}:
\begin{eqnarray}
    \Gamma_{\parallel \mathrm{chain}} &=& 4B_{1g} + 2B_{2g} + 4B_{3u} + 2A_u,\nonumber \\
    \Gamma_{\perp \mathrm{chain}} &=& 6A_g + 6B_{1u} + 6B_{2u} + 6B_{3g}. 
\end{eqnarray}
In this study, we focus on the optically active IR modes which break global inversion symmetries.
There are three types of IR modes: B$_{1u}$, B$_{2u}$, and B$_{3u}$.
Each of these IR modes break one mirror symmetry in the \emph{reciprocal} space in addition to the inversion symmetry.
In Fig. 1, we present the vibration modes of the three different types of IR modes projected on the
$\mathbf{b}-\mathbf{c}$ plane. From now on, these $\Gamma$-point IR modes are labelled depending on their symmetry and order in energy which are tabulated in Table. S1 of the Supplementary Information (SI)~\cite{Supplementary}.  


\ignore{
\begin{figure}[htb]
    \begin{center}
          \includegraphics[width=0.50\textwidth]{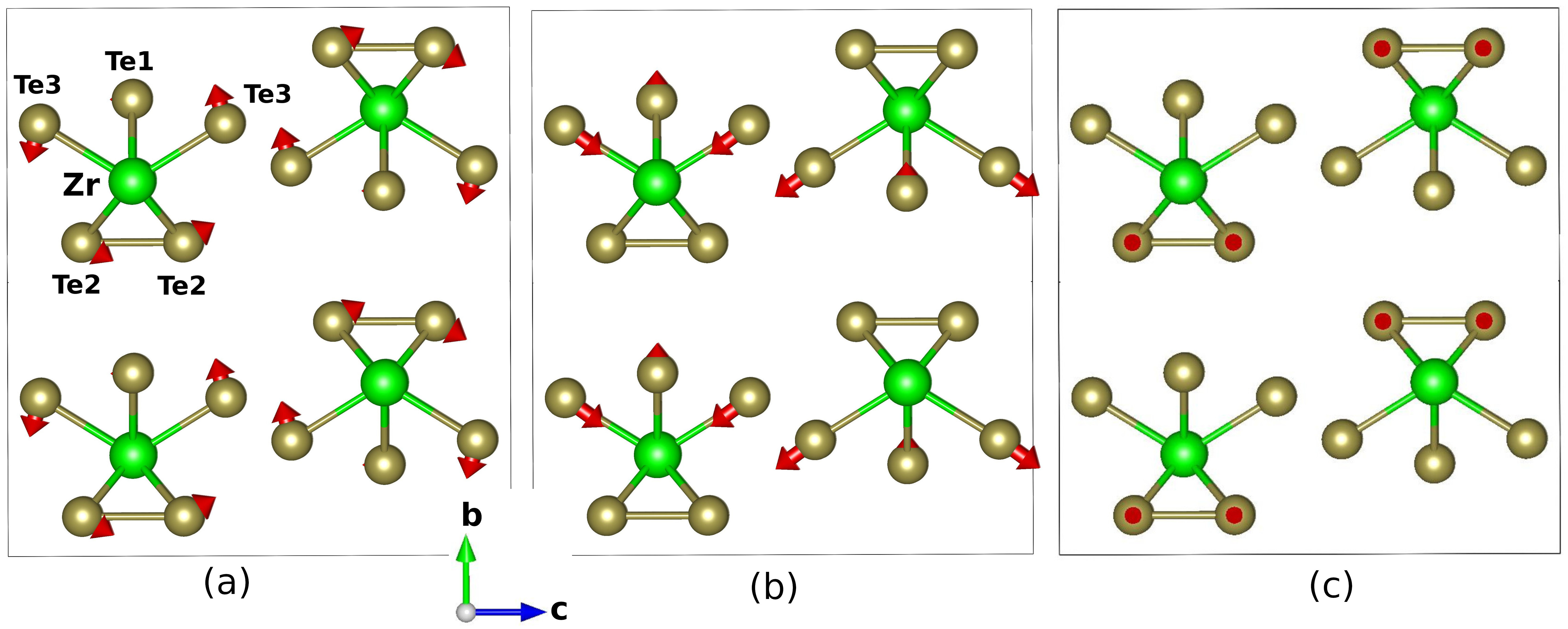}
            \end{center}
            \caption{B$_{1u}$ phonon mode and Weyl bands: (a-c) Band structure along W1-$\Gamma$-W2 direction for different $Q$ values corresponding to the B$_{1u}$-4 phonon mode. 
                    W1 and W2 are the sink and source of the Berry curvature separated along the $k_y$ direction (see panel g). The inset in Fig. (a) shows the ground state bands (i.e $Q$=0) in the vicinity of the $\Gamma$-point along with the Te3-Te2 band characters (shown in red and blue colored dots).
                    (d) Charge density plot of the homo and lumo states at the $\Gamma$ point showing the transfer of charge from Te2 to Te3 atoms for non-zero $Q$.
                    (e \& f) 2D bands forming the WPs on the $k_x$-$k_y$ ($k_a$-$k_b$) plane for different $Q$ values.
                    (g) Evolution of the WPs as a function of $Q$. $\pm1$ indicates the chirality of the WPs.
                    (h) Energy difference per formula unit as a function of $Q$ showing the harmonic regime. 
                     The blue dots in (h) highlight different $Q$ values studied in the previous figures.
                }
        \label{fig:fig1} 
\end{figure}
}

\subsection{Topological phase transition}
In this section, we present our DFT calculated electronic dispersions in the vicinity of the $\Gamma$-point 
as a function of $Q^{(l)}_{k=\Gamma}$ (written as $Q$ for convenience) which is the amplitude of oscillation (lattice distortion) of the $l^{th}$ phonon mode  with frequency $\omega^{(l)}$ at the $\Gamma$-point. 
$Q$ is expressed in units of \AA{amu}$^{\frac{1}{2}}$ such that the energy cost of the lattice distortion corresponding to the $l^{th}$ phonon mode is given by $E^{(l)}=\frac{1}{2} {\omega^{(l)}}^2 {Q^{(l)}}^2$. 
We determine the critical value of $Q$ ($Q_c$)  required 
for the formation of the Weyl points (WPs) and track their evolution (creation/annihilation and position) as a function of $Q$ for all three types of IR modes.

\subsubsection{STI-Weyl semimetal}
We find that all IR modes studied in this work drive the system from STI to the Weyl phase for $Q \sim 1$ except one IR mode which will be discussed more in the next section.
Thus formed WPs are robust over fairly large value of $Q$. 
The microscopic mechanism for the formation of the WPs is similar for all the IR modes; there is a transfer of charge from Te2 to Te3 atoms.
Also only 2 pairs of WPs are formed which is the minimum number of WPs possible for the inversion symmetry (IS) broken system.
The exact position of the WPs for different IR modes as a function of $Q$ are tabulated in SI~\cite{Supplementary}. 
The location of the WPs are consistent with our findings from the model Hamiltonian which will be discussed in the next section.
When the WPs finally gap out, the system goes to the WTI phase.
The energy and bulk band dispersion is symmetric for positive and negative values of $Q$; 
however, the chirality of the WPs and hence the sign of the Berry curvature (BC) changes 
when $Q$ value changes from positive to negative. 
This has important implications on the BC related properties which will be discussed later. 

\ignore{
\begin{figure}[htb]
    \begin{center}
            \includegraphics[width=0.50\textwidth]{fig2_new.jpg}
            \end{center}
            \caption{(a-c) Band structure and orbital content of the bands forming the WPs along the $k_z$ direction for different $Q$ values corresponding to the the B$_{2u}$-11 phonon mode: $Q$ = (a) 1 and (b) 4. 
    Fig. (c) shows bands for $Q$=1.5 starting from a DSM phase which was obtained by the application of the A$_g$-27 phonon mode ~\cite{aryal2020topological}.
                    (d) Evolution of the WPs as a function of $Q$ when starting from a DSM phase.
                    (e) Variation of the inversion symmetry breaking term, $c$ and mass, $m$ as a function of $Q$ obtained by fitting the DFT bands to the eigenvalues from the effective Hamiltonian
                        for B$_{2u}$-11 mode starting from the STI phase (i.e. the ground state) and DSM phase.
            }
            \label{fig:B2u11WPs}
\end{figure}
}

Fig. 2 shows the mechanism of formation of the Weyl bands for the lattice distortion corresponding to B$_{1u}$-4 IR mode which is also the lowest optical phonon mode with frequency of 0.58 THz.
This mode is characterized by the rotational motion of the Te atoms on each \mater~ pentamer thereby breaking $M_c$ symmetry which originally is the symmetry of the pentamer [Fig. 1(a)].
Such lattice distortion gives rise to two pairs of WPs for $|Q| \ge 1$ 
on the $k_a$-$k_b$ ($k_x$-$k_y$) plane. 
For larger $Q$ values, the WPs move away from the $k_y$ line towards the $k_x$ direction and finally annihilate for $|Q| \ge 5$ [Figs. 2(a-c) and (e,f)]. 
In addition to the increasing separation between the source and the sinks of the WPs, they also move closer to the Fermi level for higher $Q$ values.
Moreover, a nodal line forms between the valence band and the conduction band pairs.
The WPs finally annihilate for $Q \sim 6$.
The charge density plot around the \mater~ pentamer corresponding to the highest occupied (homo) and lowest unoccupied (lumo) states at the $\Gamma$-point for $Q=0$ (i.e. the ground state) and $Q=-4.5$ verifies that there is a charge transfer from Te2 to Te3 atoms mediated by the Zr atoms during this dynamical evolution of the atoms [Fig. 2(d)]. 
In Fig. 2(g), we plot the position of the sources and sinks of the WPs for different $Q$ values on the $k_x$-$k_y$ plane
 along with their chiralities which show the evolution of the WPs (i.e. creation, movement and annihilation). 
Such dynamical manipulation of the WPs in momentum space could find applications for braiding purposes~\cite{trevisan2021bicircular} especially if the paths of the different Weyl nodes  interchange. 
This might be possible in the presence of other time dependent perturbations or if more mirror symmetries are broken in the system dynamically, e.g. by the application of other IR modes.
Fig. 2(h) shows the energy cost per formula unit as a function of the lattice distortion corresponding to the B$_{1u}$-4 mode (expressed in units of $Q$) which verifies that the system is indeed within the harmonic regime during this dynamical evolution.
Similar band plots and analyses for B$_{3u}$-8 and B$_{2u}$-20 IR modes are presented in the SI. 

\subsubsection{WTI-Weyl semimetal}
We also studied the possibility of the IR phonon mode induced topological phase transition starting from the WTI phase.
The WTI phase was obtained from the application of the A$_g$-27 phonon mode\cite{aryal2020topological}.
We find that for the case of the B$_{1u}$ mode, the system does not undergo transition to the Weyl phase irrespective of the proximity of the WTI system to the WTI-STI phase boundary.
However, for the case of the B$_{3u}$ mode, such transition is possible if the system is close to the phase bounday.
Both observations are consistent with the $\mathbf{k} \cdot \mathbf{p}$ model (Eq. \ref{eqn:H_kp}).
In the absence of the $\mathbf{k}$-dependent perturbation term, the topological phase transition from the WTI to the Weyl phase is not possible for all the IR phonon modes.
However, the presence of the $\mathbf{k}$-dependent perturbation term allows such transition (see SI~\cite{Supplementary} for details).

\subsubsection{DSM-Weyl semimetal}
The critical value of $Q$  required for the formation of the WPs  ($Q_c$) can be decreased if one starts from DSM phase instead of the gapped phase.
This is best demonstrated in Fig.~3 with the case of B$_{2u}$-11 mode, where we see a drastic variation in $Q_c$: for a gapped STI phase, the system enters the Weyl phase for $Q>6$ which corresponds to the energy cost of more than 100 meV/f.u. [Figs. 3(a) and 3(b)]. However, 
if the initial phase is a DSM, which can be easily obtained by the application of resonant A$_g$ Raman modes (the A$_g$-27 phonon mode was used here~\cite{aryal2020topological}, then the system enters the Weyl phase for an infinitesimal value of $Q$ [Figs. 3(c), 3(d) and 3(f)].
Though not as dramatic, this is also true for the case of B$_{1u}$-4 mode.

This implies that in pump-probe experiments, a low-power resonant IR light source can drive the system to the Weyl phase if one first prepares the sample to be close to a DSM phase by static perturbations such as strain~\cite{StrainTunedTopology_Mutch_Science2019}. Another exciting approach is to follow a photon pumping that excites resonant A$_g$ modes to induce a DSM phase~\cite{aryal2020topological,Konstantinova2020photoinduced,RamanZrTe5_VaswaniPRX2020}, leading to an all-light-controlled method, as illustrated in Fig.~3(g). 

\subsection{Effective Hamiltonian for IR modes}
In order to understand the process of formation of the WPs for all possible IR phonon modes of \mater, we used the extended $\mathbf{k} \cdot \mathbf{p}$ model of Chen \textit{et. al.}  ~\cite{ChenEffectiveH_PRL_2015,ZrTe5_Choi_Kyungwha_PRB2020} to devise an effective Hamiltonian describing the essential low energy physics  of the STI phase of ~\mater~ near the $\Gamma$-point. The Hamiltonian without inversion symmetry breaking perturbations is: 
\begin{equation}
        H(\mathbf{k}) = (m-Dk^2)\tau^z + v_xk_x\tau^x\sigma^y + v_yk_y\tau^x\sigma^x + v_zk_z\tau^y,
    \label{eqn:H_kp}
\end{equation}
where, $Dk^2=D_1k_x^2+D_2k_y^2+D_3k_z^2$.
Here the $\sigma$ Pauli matrices act on spin components and $\tau$ Pauli matrices act on valley indices. 
 The low energy Hamiltonian is derived by enforcing the two mirror symmetries perpendicular to the crystal \textbf{c}-axis and \textbf{a}-axis, $M_c$ and $M_a$ respectively, time reversal symmetry (TRS) and inversion symmetry (IS) where:
 $M_{c}=\tau^z\cdot i\sigma^z$,
 $M_{a}=i\sigma^x$
 $T=K \cdot i\sigma^y$ and 
 $I=\tau^z$.
 Because of the presence of the  two mirror symmetries and inversion symmetries, the Hamiltonian also has a third symmetry, $G_b$, a glide symmetry (mirror plus translation parallel to the mirror plane) perpendicular to the \textbf{b}-axis.

The IR phonon modes produce two major effects on (\ref{eqn:H_kp}): they change $m$ and add  inversion symmetry breaking perturbations.
Depending on how the IS is broken, the mirror symmetries are also broken differently for each IR modes.
For example, all B$_{1u}$ type IR modes break $M_c$
whereas B$_{2u}$ (B$_{3u}$) modes break $G_b$ ($M_a$).

From simple symmetry analysis, it is straightforward to obtain the leading order terms for each of the IR modes. 
For example, for B$_{1u}$ mode, 
the {\bf {k}}-independent leading order perturbation to Hamiltonian (\ref{eqn:H_kp}) is just $c\tau^x$ where $c$ is a constant.
The eigenvalues for this case are given by:
\begin{eqnarray}
E^{r}_{s}=r\sqrt{(m - Dk^2)^2 + \tilde{k}_z^2 + (|\tilde{k}_{xy}|s+ c)^2},
\label{eqn:evals_B1u4}
\end{eqnarray}
where, $r,s \in \pm$. 
The superscript $r$ denotes the valence and conduction states and the subscript $s$ denotes the states within the same branch.
Two bands $E^-_+$ and $E^+_-$ cross at zero energy for $k_z=0$ only if two ellipses of the form:
\begin{eqnarray}
D_1k_x^2 +D_2k_y^2 = m, 
~v_x^2k_x^2 +v_y^2k_y^2 = c^2
\label{Eq:ellipses}
\end{eqnarray}
intersect at finite number of $\bf{k}$ values.
Thus formed band touching points are WPs with  the linear dispersion in the vicinity of these band touching points as shown in the SI~\cite{Supplementary}. 
Using the values of $D$'s, $v$'s and $m$  obtained from the fits to the ground state solution (i.e. for $Q$ =0) ~\cite{aryal2020topological},
we find that the two ellipses intersect for $c > 0.01$ thereby giving 4 WPs on the $k_x-k_y$ plane (see SI).

The symmetry considerations dictate that {\bf k}-dependent  perturbation term  for the B$_{1u}$ IR modes can only be of the form $(\alpha k_x\sigma^y + \beta k_y\sigma^x)\tau^z$, where $\alpha$ and $\beta$ are constants.
Such term does not create a gap, it only moves the WPs on the $k_x-k_y$ plane. One can perform a similar analyses for other IR modes.
The $k$-independent and dependent perturbation terms for all three types of IR modes are shown in Table ~\ref{table:ISBreakingTerms}]. 
In the absence of the  {\bf {k}}-dependent perturbation, 
the eigenvalues can be found analytically.
DFT calculation for the corresponding phonon modes gives the WPs on the same plane found from this simple analysis.

The magnitude as well as variation of the leading order inversion symmetry (IS) breaking perturbation term in the model Hamiltonian ($c$-value) can be obtained 
by fitting the DFT eigenvalues in close vicinity of the  $\Gamma$-point with the analytical results [Table ~\ref{table:ISBreakingTerms}] as a function of $Q$.
Fig. 3(e) shows the variation of $c$ and mass $m$ as a function of $Q$ for B$_{2u}$-11 mode when perturbing the system from the gapped as well as the DSM phase.
As expected, $c$ varies almost identically for both cases whereas the $m$ curves have similar behavior but shifted from one-another due to the zero mass of the DSM phase.
The variation of $c$ and mass $m$ as a function of $Q$ for B$_{1u}$-4 and B$_{3u}$-8 are shown in the SI~\cite{Supplementary}. 
\subsection{Nonlinear Berry curvature effect}
It is well known that for systems like \mater~ with both IS and time reversal symmetry (TRS),
Berry curvature ${\bf \Omega}({\bf k})$ is identically zero everywhere in the Brillouin zone~\cite{BerryPhaseRMP_Xiao_Niu2010}.
However, when IS is broken as is the case here, 
${\bf \Omega}({\bf k})$ is finite and contributes to different transport phenomena.
Here, we focus on the second order effects arising from the Berry curvature such as  photo-galvanic effects and 
nonlinear anomalous Hall effect (NLAHE). The latter one,  
unlike the linear AHE,  does not require broken time reversal symmetry~\cite{QNLHE_Sodemann_Fu_PRL2015}.

The intrinsic contribution (intra-band) to the afore-mentioned nonlinear effects
can be understood in terms of the Berry curvature dipole moment (BCDM).
BCDM is dimensionless in 3D and is zero for a purely isotropic Weyl cone~\cite{QNLHE_Sodemann_Fu_PRL2015} [see Methods section for definition of BCDM]. 
In \mater, because of the presence of the different mirror symmetries, different components of the BCDM are constrained to be zero by symmetry.
For example, for B$_{1u}$ mode, the presence of the $M_a$ and $G_b$ symmetry dictates the following transformation rules for the velocities and the BCDM tensor:
\begin{equation}
\begin{gathered}
(v_x,v_y,v_z) \xrightarrow{M_a} (-v_x,v_y,v_z), 
(v_x,v_y,v_z) \xrightarrow{G_b} (v_x,-v_y,v_z),\\
(\Omega_x,\Omega_y,\Omega_z) \xrightarrow{M_a} (\Omega_x,-\Omega_y,-\Omega_z),\\ 
(\Omega_x,\Omega_y,\Omega_z) \xrightarrow{G_b} (-\Omega_x,\Omega_y,-\Omega_z).
\end{gathered}
\label{eqn:BCDM_Ma_Gb_symm}
\end{equation}
Hence, $M_a$ symmetry constrains all other terms of BCDM to vanish except $D_{xy}, D_{yx}, D_{xz}, D_{zx}$.
Similarly, from $G_b$ symmetry, we find that only $D_{xy}, D_{yx}, D_{yz}, D_{zy}$ survive.
Enforcing both $M_a$ and $G_b$ symmetries, only $D_{xy}$ and $D_{yx}$ terms survive for B$_{1u}$ type IR modes. 
Using similar arguments, we find that for B$_{2u}$ mode, only $D_{xz}$ and $D_{zx}$ terms survive whereas for B$_{3u}$, $D_{yz}$ and $D_{zy}$ terms survive.
This will have nonlinear response along different directions. 
In the following, we present the results for the BCDM in B$_{1u}$-4 mode only as its magnitude is larger compared to other modes.
This is a direct consequence of the symmetry of the phonon modes and anisotropic band dispersion.

BCDM, being a Fermi surface (FS) property, depends on the shape and size of the FS, the vicinity of the WPs from the FS \textit{etc}. 
Fig. 4(a-c) shows the evolution of the FS for different $Q$ values corresponding to the B$_{1u}$ mode lattice distortion.
The FS forms a small hole pocket in the vicinity of the $\Gamma$-point however 
its topology changes drastically during this evolution. 
Moreover, the magnitude of $\mathbf{\Omega}$ increases during this evolution
because the Fermi level shifts closer to the WPs.
The magnitude of $\Omega_y$ is about 5 times larger compared to $\Omega_x$ and $\Omega_z$ (see SI).
Since both of the factors, $v_x$ and $\Omega_y$ that appear in the evaluation of $D_{xy}$ are bigger compared to that of 
$D_{yx}$ which involves $v_y$ and $\Omega_x$, $D_{yx}$ is negligible compared to $D_{xy}$. 

Fig. 4(d) shows the variation of $D_{xy}$ with the chemical potential ($\mu$) obtained from the \textit{ab-initio} calculation. The value of $\mu=0$ corresponds to the charge neutral point.
The average peak value of $D_{xy}$ ($\sim$ 0.05) at $\mu \sim 10$ meV is similar to that of another type-I Weyl semimetal TaP and presumed type-II Weyl semimetal WTe$_2$~\cite{BCDM_Zhang_Yan_PRB2018}.
As expected, the peak value of BCDM  is concentrated around the WP energy (denoted by the black vertical line in Fig. (d)). 
When $\mu$ is right at the WP energy (for $|Q| >1$) or in the bandgap (for $|Q| <1$), BCDM vanishes due to the vanishing FS.
$D_{xy}$ possesses a striking  feature which distinguishes the Weyl from the non-Weyl phase: it changes sign as a function of $\mu$ only when the system hosts WPs.
The reason for such sign change after crossing the WP energy is simple to understand.
It is obvious that the sign of $\mathbf{\Omega}$ is different for the valence and conduction bands\cite{CommentBerryCurvature}.
This is also true for band velocity $\mathbf{v}$ when Weyl cone does not have a tilt.
However, due to the presence of the $\mathbf{k}$-dependent constant terms, the tilt is finite here and the sign of $v_x$ stays the same.
Hence, $D_{xy}$ changes sign upon crossing the WP energy.

In Fig. 4(e), we show the variation of BCDM as a function of $Q$ for different values of $\mu$.
The asymmetry around $Q=0$ is due to the switching of the chirality between the negative and positive $Q$ values and is present for all values of $\mu$.
When $\mu$ is positioned at the conduction band, in addition to the sign flip at $Q=0$, D$_{xy}$ changes sign when the system enters the Weyl phase at around $Q$ of 1.5.
Hence, depending on the position of the Fermi level, which can be tuned by doping or even temperature~\cite{LifshitzZrTe5_Zhang_NatureComm2017,ZrTe5Lifshitz_QLi_IOP2017}.
BCDM and the associated currents change sign multiple times as a function of $Q$.
\section{Discussion} 
Since only $D_{xy}$ and $D_{yx}$ components of the BCDM are non-zero for B$_{1u}$ mode, 
from Eq.~\ref{eqn:conductivity}, we find that only non-vanishing components of the conductivity tensor are 
$\sigma_{xxz} = - \sigma_{zxx}$ from $D_{xy}$ and 
$\sigma_{zyy}= - \sigma_{yzz}$ from $D_{yx}$.
As $D_{yx}$ is negligible compared to $D_{xy}$, we focus on the $\sigma$ component arising from $D_{xy}$.
Since the nonlinear current is given by 
$j_a = \sigma_{abc} E_b(\omega)E_c^*(\omega)$,
$\mathbf{E}(\omega)$ being the applied electric field,
$\sigma_{zxx}$ component dictates that application of field along the $\mathbf{x}$-direction (i.e. $b=c=x$) produces a current along the $\mathbf{z}$-direction
which amounts to the nonlinear Hall effect in the limit of $\omega \rightarrow 0$.
Fig. 4(f) shows schematic diagram of the non-linear Hall response in the presence of an external transverse electric field and IR photon in resonance with the  B$_{1u}$ phonon mode.
Considering the maximum value of $D_{xy}$ of $\sim$ 0.1 and relatively longer relaxation time for the chiral particles $\tau$ $\sim 10 ps$~\cite{Luo_NP_21_ZrTe5}, we find that a maximum non-linear Hall current ($J_z$) of about 2 $\frac{nA}{\mu m^2}$ can be generated in the presence of a typical laboratory electric field ($E_x$) $\sim$ 100 $\frac{V}{m}$ and IR photon radiation in resonance with the B$_{1u}$-4 phonon mode.
We note that our predictions regarding BCDM induced non-linear anomalous hall currents for different IR phonon modes are also applicable to other static IS breaking perturbations, e.g. in the presence of a strong electric field.
In that case, the external perturbations breaking IS correspond to the linear superposition of the different IR phonon modes which will give rise to BCDM induced Hall currents in different directions.


Finally,  we would also like to briefly comment  on the photocurrent experiment discussed in Ref.~\onlinecite{Luo_NP_21_ZrTe5}.
It is mentioned there that the photocurrent measured by the THz irradiation via circular photogalvanic effect (CPGE) along the crystal $\mathbf{a}$-axis ($\mathbf{x}$-direction) 
is much larger than that measured along the crystal $\mathbf{c}$-axis ($\mathbf{z}$-direction)
given that the ratio of the terahertz emission polarization along the $\mathbf{z}$ and $\mathbf{x}$-direction ($E_z$:$E_x$)is $9:1$.
If the dominant IR phonon mode induced by the terahertz pump is indeed B$_{1u}$-4 as claimed, then such asymmetry can be understood from the above analysis. 
The currents along the $\mathbf{x}$ and $\mathbf{z}$-directions are given by
$j_x = \sigma_{xxz} E_x(\omega)E_z^*(\omega)$ 
 and $j_z = \sigma_{zxx} E_x(\omega)E_x^*(\omega)$ respectively.
Hence, it is easy to see that $j_x$ will be significantly higher (by 9 times) than $j_z$ as $\sigma_{xxz} = - \sigma_{zxx}$.

\ignore{
We predict that for larger value of $Q$ when the system achieves the Weyl phase by coupling with phonons as it does for the current case (or by some other means), 
the Hall current switches its direction.
Namely, for the specific example of B$_{1u}$-4 mode, when $Q <1$, the nonlinear Hall current is positive, whereas it is negative for $Q\ge 1$.
Thus, the system acts as a switch which operates on ultrafast time scale.
}

\subsection{Conclusion and Outlook}
\label{Sec:Conclusion}
In summary, we find that atomic displacements corresponding to any of the three types of infrared modes of \mater~ 
can drive the system from small bandgap topological insulating phase to a Weyl semimetallic phase by breaking the global crystal inversion symmetry in one of the three possible ways.
The position of the WPs are constrained by the mirror symmetries, 
and the WPs remain robust over a large value of the atomic displacements corresponding to the phonon modes.
The magnitude of the atomic displacements necessary for driving the system into the Weyl phase can be reduced dramatically if the system is a Dirac semimetallic phase. 
This can be tested in future experiments by using a relatively lower power resonant IR laser source in conjunction with A$_g$ Raman mode pumps or applying resonant IR laser source to strained ~\mater~in close vicinity to the Dirac semimetallic phase.
We also find that for the lowest optical phonon mode, the sign of the BCDM, which gives rise to various nonlinear effects,  
changes upon crossing the Fermi level for only Weyl phase suggesting an experimental way to detect photo-induced Weyl phase in ~\mater.

In condensed matter physics, geometrical chirality of a crystalline electronic system is normally fixed by the chiral lattice structure of a material on formation, when it lacks mirror planes, space-inversion centers, or rotoinversion axes. 
Dynamic chirality generation by inversion-symmetry breaking paves the way for the development of disorder-tolerant quantum electronics through electromagnetic topology control.
While first-principles dynamical simulations 
are needed to make a more direct comparison with the ultrafast experiments, the present work sheds light on this research direction.

\section{Methods}
First principles density-functional-theory (DFT) calculations were done using Quantum Espresso (QE)~\cite{QE-2009} package. 
Perdew-Burke-Ernzerhof (PBE) exchange-correlation functional~\cite{PBE} within the generalized gradient approximation (GGA) were used in all the calculations.
Full lattice relaxation and subsequent phonon and electronic band calculations using both QE and VASP packages~\cite{VASP1,PAW}.
For QE, we used  fully relativistic norm conserving pseudopotentials generated using the optimized norm-conserving Vanderbilt pseudopotentials~\cite{ONCVPPHamann2013}. 
The lattice relaxation was performed using Grimme's semi-empirical DFT-D3 vdW interaction as implemented in the QE software as it gave the best agreement with the experimental lattice parameters (with just 3 $\%$ deviation in volume). 
The primitive BZ was sampled by using k mesh of 10 $\times$ 10 $\times$ 8 and energy cutoff of 1000 eV was used after careful convergence tests. 
For VASP, the optB86b vdW correlation functional was adopted to account for vdW interactions. 
Density functional perturbation theory (DFPT) method was used to compute the atomic forces as implemented in Phonopy~\cite{Phonopy} under the harmonic approximation.

Wannierization method without localization~\cite{Wannier902014} was employed to extract tight-binding Hamiltonian in the basis of all Zr-$d$ and Te-$p$ orbitals which was subsequently used to find the location of the Weyl nodes by using Wannier Tools package~\cite{WannierTools}. 
Berry curvature and  related properties were calculated following the Kubo formula as implemented in the Wannier90 package~\cite{Gyrotropiceffects_Tsirkin_Souza_PRB2018}.

The third rank conductivity tensor associated with the BC effect is defined within the relaxation time approximation as~\cite{QNLHE_Sodemann_Fu_PRL2015}:
\begin{equation}
    \sigma_{abc} = -\frac{e^3\tau}{2\hbar^2(1+i\omega\tau)}\epsilon_{adc}D_{bd},
    \label{eqn:conductivity}
\end{equation}
where, $\tau$ is the relaxation time and $D_{bd}$, the Berry curvature dipole moment (BCDM) tensor, is in general a function of the chemical potential $\mu$  and  is given by:
\begin{eqnarray}
    D_{ab} (\mu) = \int \frac{d^3 {\bf k}}{(2\pi)^3}\sum_n v_a^n({\bf k}) \Omega_b^n({\bf k}) \Big(\frac{\partial f_0(E,\mu)}{\partial E}\Big)_{E=E_{kn}}.
   \label{eqn:BCDM}
\end{eqnarray}
$v_b^n({\bf k})$ is the $b^{th}$ component group velocity of the $n^{th}$ band given by $\frac{\partial E_{nk}}{\partial k_b}$, 
$f_0$ is the equilibrium occupation factor and $\Omega_{bc}^n({\bf k}) = \epsilon_{abc}\Omega_{a}^n({\bf k})$ is given by:
\begin{eqnarray}
    \Omega^n_{ab}(\mathbf{k}) = -2\sum_{m \neq n} \mathrm{Im} \frac{\langle n\mathbf{k} | \hat{v}_a|m\mathbf{k}\rangle \langle m\mathbf{k} | \hat{v}_b|n\mathbf{k}\rangle}{(E_{nk} - E_{mk})^2},
    \label{eqn:BerryCurvature}
\end{eqnarray}
For the calculation of the BCDM, 
a very dense k-point mesh (upto $300\times 300\times 300$) was used to sample a small volume around the $\Gamma$-point in order to capture the rapidly varying distribution of the Berry curvature around the regions of band crossings. 
Smearing of 5 meV was used.

The parameters of the effective Hamiltonian (~\ref{eqn:H_kp}) were extracted by fitting the analytical eigen values with the  DFT eigen values in close vicinity of the $\Gamma$ point (upto 0.015 \AA$^{-1}$).
We find that as a function of $Q$, other parameters of the effective Hamiltonian like $D_i$ and $v_i$ does not vary much as compared to the $c$ and $m$ parameters which is the main effect of the IR phonon modes.

\section{Data Availability}
The data that support the findings of this study are available from the corresponding author upon reasonable request.

\section{Code Availability}
Quantum espresso, Wannier90, WannierTools, and VESTA programs used in this work are publicly available.

\section{Acknowledgements}
This work was supported by U.S. Department of Energy (DOE) the Office of Basic Energy Sciences, Materials Sciences and Engineering Division under Contract No. DE-SC0012704. X.J. acknowledges the visiting scholarship of Brookhaven National Laboratory and the financial support of China Scholarship Council
and National Natural Science Foundation of China (No. 11774119).

\section{Author Contributions}
W.Y., N.A., and Q.L. designed the project. N.A. performed the first-principles electronic band structure calculations and effective Hamiltonian analysis. X.J. performed the full lattice relaxation and phonon calculations. A.T. provided theoretical insights. All authors contributed to the analysis and discussion of the results. N.A. and W.Y. wrote the manuscript with input from all coauthors.

\section{COMPETING INTERESTS}
The authors declare no competing interests.

\section{ADDITIONAL INFORMATION}
\noindent \textbf{Correspondence} and requests for materials should be addressed to Niraj Aryal and Weiguo Yin.

\bibliography{references}

\newpage

\onecolumngrid
\section{Figure legends}
Fig. 1: \textbf{Three different types of IR phonon modes of \mater~ projected onto the $\mathbf{b}-\mathbf{c}$ plane with the vectors showing the normal atomic displacement.} (a) B$_{1u}$-4 (b) B$_{2u}$-20 and (c) B$_{3u}$-8 modes with frequencies of 0.58, 3.04 and 1.38 THz respectively. 
The atomic displacements in panel (c) are perpendicular to the $\mathbf{b}-\mathbf{c}$ plane, 
hence the vectors are not seen in the figure. 
\\

Fig. 2: \textbf{B$_{1u}$ phonon mode and Weyl bands.} (a-c) Band structure along W1-$\Gamma$-W2 direction for different $Q$ values corresponding to the B$_{1u}$-4 phonon mode. 
W1 and W2 are the sink and source of the Berry curvature separated along the $k_y$ direction (see panel g). The inset in Fig. (a) shows the ground state bands (i.e $Q$=0) in the vicinity of the $\Gamma$-point along with the Te3-Te2 band characters (shown in red and blue colored dots).
(d) Charge density plot of the homo and lumo states at the $\Gamma$ point showing the transfer of charge from Te2 to Te3 atoms for non-zero $Q$.
(e \& f) 2D bands forming the WPs on the $k_x$-$k_y$ ($k_a$-$k_b$) plane for different $Q$ values.
(g) Evolution of the WPs as a function of $Q$. $\pm1$ indicates the chirality of the WPs.
(h) Energy difference per formula unit as a function of $Q$ showing the harmonic regime. 
 The blue dots in (h) highlight different $Q$ values studied in the previous figures.
\\

Fig. 3: \textbf{Weyl bands formation in B$_{2u}$-11 mode.} (a-c) Band structure and orbital content of the bands forming the WPs along the $k_z$ direction for: $Q$ = (a) 1 and (b) 4. 
Fig. (c) shows bands for $Q$=1.5 starting from a DSM phase which was obtained by the application of the A$_g$-27 phonon mode ~\cite{aryal2020topological}.
(d) Evolution of the WPs as a function of $Q$ when starting from a DSM phase.
(e) Variation of the inversion symmetry breaking term, $c$ and mass, $m$ as a function of $Q$ obtained by fitting the DFT bands to the eigenvalues from the effective Hamiltonian for B$_{2u}$-11 mode starting from the STI phase (i.e. the ground state) and DSM phase. (f) Momentum transfer between the Weyl points as a function of $Q$ for pumping a STI (blue dots) and a DSM (red dots). (g) Schematics of a double-pumping all-light-controlled experiment that utilizes low-power laser sources to drive ZrTe$_5$ into a Weyl state.
\\

Fig. 4: \textbf{Berry curvature (BC) and related transport signatures in B$_{1u}$ mode.} (a-c) Intensity plot showing the distribution of the $\Omega_y$ component of the BC on the FS calculated from the model Hamiltonian for different values of $c$ which roughly corresponds to $Q$ of 1, 2 and 4 respectively. The Fermi level is set from the DFT results.
(d \& e) D$_{xy}$ component of BCDM as a function of $\mu$ and $Q$ respectively calculated using the \textit{ab-initio} Hamiltonian. $\mu=0$ corresponds to the charge neutral point.
The black horizontal line in Fig.(d) corresponds approximately to WP energy for the case when Weyl points occur whereas for the gapped case, it corresponds to the average of the valence band maximum and conduction band minimum.
(f) Schematic diagram of the non-linear transverse hall response in the presence of an external electric field and inversion symmetry breaking perturbation like B$_{1u}$ phonon mode.

\newpage
\section{Table}
\begin{table}[!tht]
\caption{Inversion symmetry breaking perturbation term for each of the IR modes added to the effective Hamiltonian.
$M=(m-Dk^2)^2$, $\tilde{k_i} = v_i k_i$
and $|\tilde{k}_{ij}| = \sqrt{\tilde{k}_i^2 + \tilde{k}_j^2}$.
}
\begin{center}
\begin{tabular}{|p{40pt}|p{50pt}|p{50pt}|p{105pt}|p{85pt}|p{60pt}|} \hline
    \multicolumn{6}{|c|}{\textbf{Model H}} \\ \hline 
 IR mode & Broken symmetry & Constant term &  Eigenvalues ($E^2$)  & {\bf{k}}-dependent term & WPs location \\ \hline
 B$_{1u}$  & $M_c$ & $c\tau^x$  &  $M^2 + \tilde{k}_z^2 + (|\tilde{k}_{xy}| \pm c)^2$ & $(k_x\sigma^y + k_y\sigma^x)\tau^z$  & $k_x$-$k_y$ plane\\ \hline
 B$_{2u}$ &  $G_b$ & $c\tau^y\sigma^x$  & $M^2 + \tilde{k}_y^2 + (|\tilde{k}_{xz}| \pm c)^2$ & $(k_z\sigma^x + k_x\sigma^z)\tau^z $ & $k_x$-$k_z$ plane \\ \hline
 B$_{3u}$ &  $M_a$ & $c\tau^y\sigma^y$  & $M^2 + \tilde{k}_x^2 + (|\tilde{k}_{yz}| \pm c)^2$  & $(k_y\sigma^z+k_z\sigma^y)\tau^z$ & $k_y$-$k_z$ plane\\ \hline
\end{tabular}
\end{center}
\label{table:ISBreakingTerms}
\end{table}

\begin{figure*}[htb]
    \begin{center}
            \includegraphics[width=0.98\textwidth]{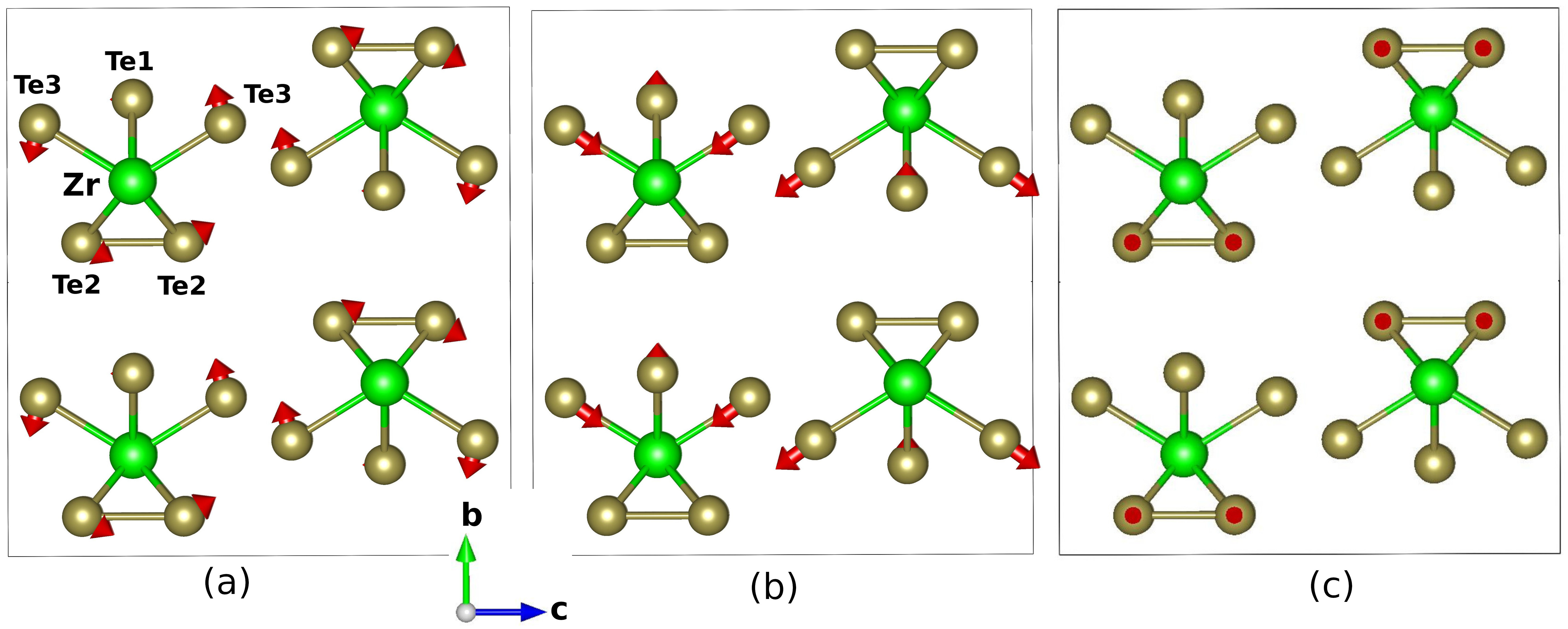}
            \caption{}
    \end{center}
        \label{fig:fig1} 
\end{figure*}

\begin{figure*}[htb]
    \begin{center}
          \includegraphics[width=0.98\textwidth]{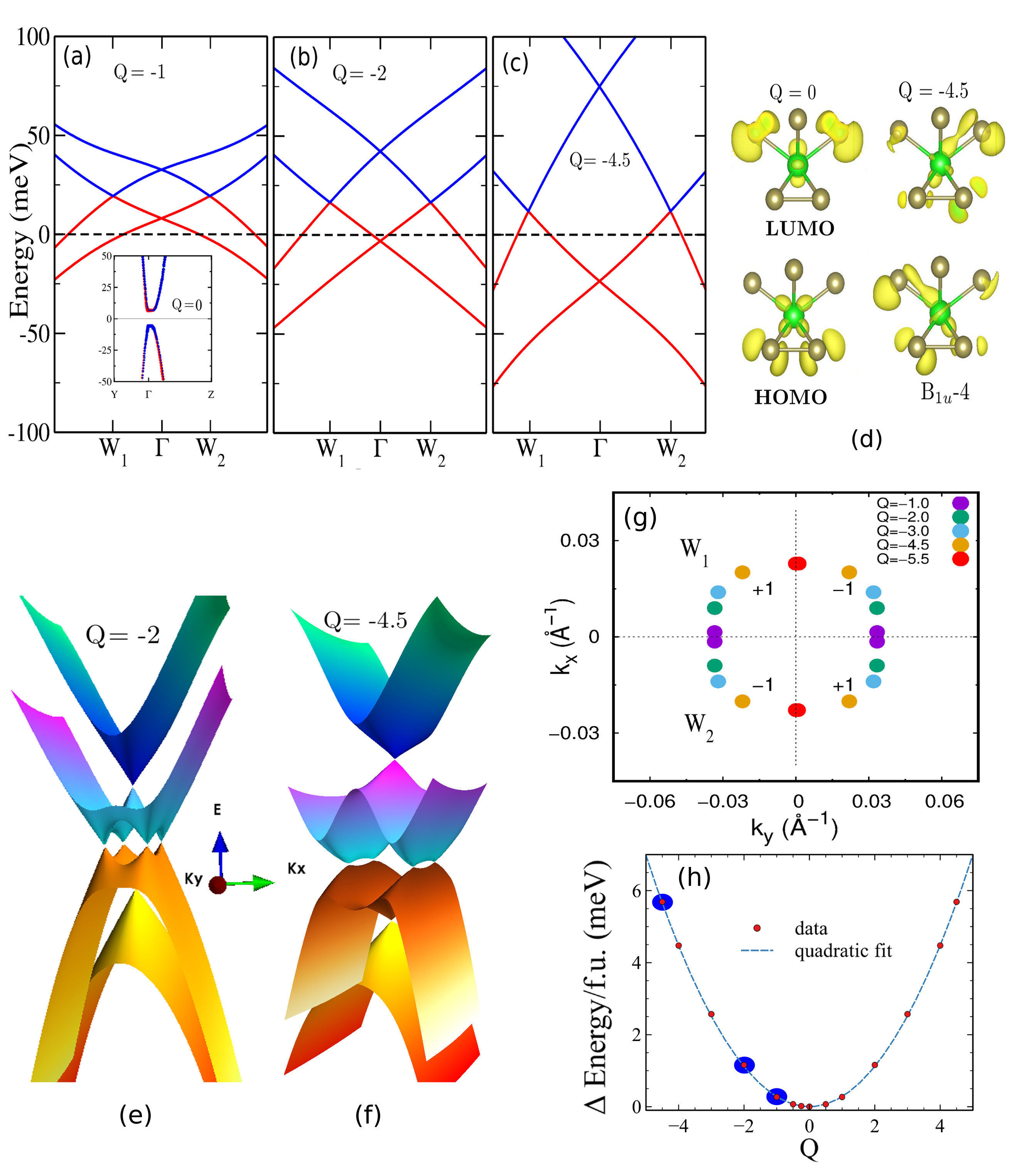}
            \caption{}
            \end{center}
        \label{fig:fig2} 
\end{figure*}

\begin{figure*}[htb]
    \begin{center}
            \includegraphics[width=0.85\textwidth]{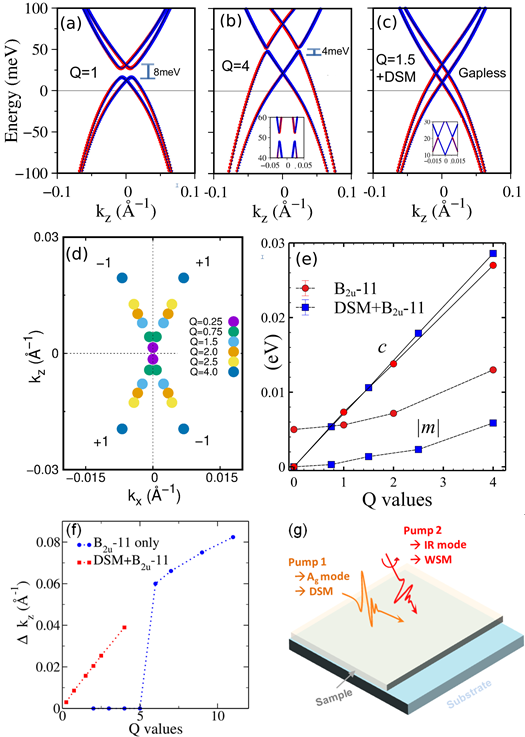}
                \caption{}
            \end{center}
            \label{fig:fig3}
\end{figure*}

\begin{figure*}[htb]
    \begin{center}
            \includegraphics[width=0.98\textwidth]{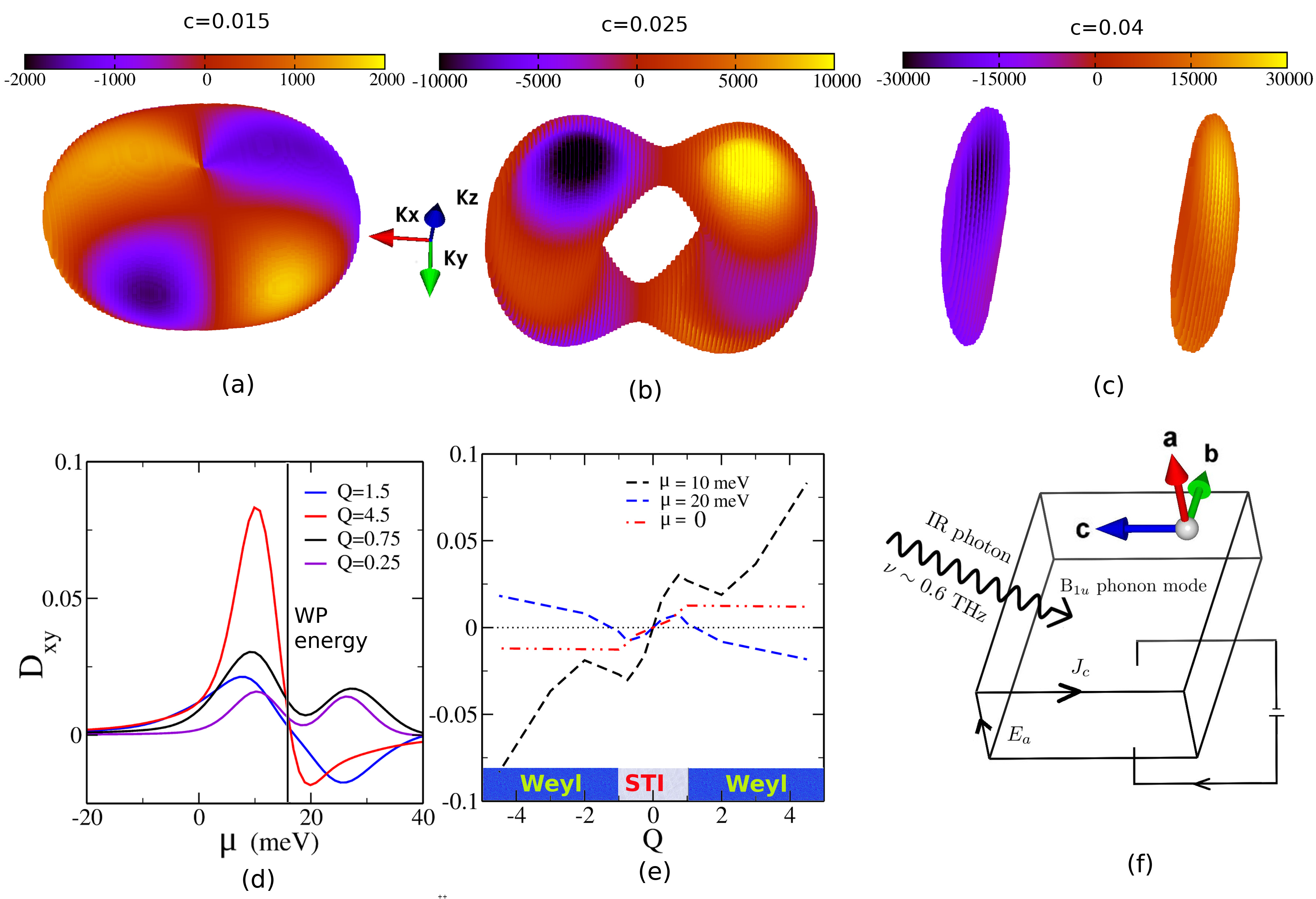}
            \caption{}
            \end{center}
        \label{fig:fig4} 
\end{figure*}

\end{document}